\documentclass{jfm}
\usepackage{graphicx}
\usepackage{epstopdf, epsfig}
\graphicspath{ {images/} }

\shorttitle{A probabilistic protocol for the assessment of transition and control}
\shortauthor{A. Pershin, C. Beaume and S.M.\ Tobias}

\title{A probabilistic protocol for the assessment of transition and control}

\author{Anton Pershin\aff{1}
  \corresp{\email{mmap@leeds.ac.uk}},
  C\'edric Beaume\aff{1}
 \and Steven M. Tobias\aff{1}}

\affiliation{\aff{1}School of Mathematics, University of Leeds, Leeds LS2 9JT, UK}

\begin{document}

\maketitle

\begin{abstract}
Transition to turbulence dramatically alters the properties of fluid flows.
In most canonical shear flows, the laminar flow is linearly stable and a finite-amplitude perturbation is necessary to trigger transition.
Controlling transition to turbulence is achieved via the broadening or narrowing of the basin of attraction of the laminar flow.
In this paper, a novel methodology to assess the robustness of the laminar flow and the efficiency of control strategies is introduced.
It relies on the statistical sampling of the phase space neighborhood around the laminar flow in order to assess the transition probability of perturbations as a function of their energy.
This approach is applied to a canonical flow (plane Couette flow) and
provides invaluable insight: in the presence of the chosen control, transition is significantly suppressed whereas plausible scalar indicators of the nonlinear stability of the flow, such as the edge state energy, do not provide conclusive predictions.
The methodology presented here in the context of transition to turbulence is applicable to any nonlinear system displaying finite-amplitude instability.
\end{abstract}

\begin{keywords}
Authors should not enter keywords on the manuscript, as these must be chosen by the author during the online submission process and will then be added during the typesetting process (see http://journals.cambridge.org/data/\linebreak[3]relatedlink/jfm-\linebreak[3]keywords.pdf for the full list)
\end{keywords}

\section{Introduction}

Controlling transition to turbulence is vitally important in a variety of applications ranging from pipeline flows to mixing devices.
In the former, turbulence undesirably enhances transport losses, whilst in the latter, it improves mixing efficiency.
A large body of research has been devoted to specific passive and active control strategies to decrease transport losses by reducing the friction drag of turbulent flows \citep{Jung1992, Baron1995, Quadrio2004, Kasagi2009, Quadrio2011, Brunton2015}.
Another ongoing effort to control transition is aimed at manipulating the robustness of the laminar flow to perturbations via passive modifications of channel surfaces and linear feedback control techniques \citep{Sreenivasan1982, Kim2007, Hof2010, Rabin2014, Marensi2019}.
The resulting strategies are largely based on the knowledge that, in subcritical shear flows, transition to long-lived turbulence occurs when both the Reynolds number, quantifying the ratio of the inertia of the flow to the viscous forces, exceeds a critical value $Re_c$ \citep{Avila2011, Shi2013}, and a sufficiently energetic disturbance is applied \citep{Schmiegel1997, Duguet2013}.
A strategy designed to prevent transition can be deemed successful when it increases the volume of the basin of attraction of the laminar flow relative to the state space volume.
This relative volume is related to the \textit{edge of chaos}, the manifold separating initial conditions that decay from those that transition \citep{Skufca2006, Schneider2008, Chantry2014}.
The edge of chaos comprises local attractors, the \textit{edge states}, and their stable manifolds.
The importance of edge states has indeed been highlighted for boundary layer and pipe flows, where their neighborhood is often visited before transition \citep{Mellibovsky2009, Khapko2016}, and for plane Couette flow, where they have been related to optimal energy growth disturbances \citep{Duguet2010, Olvera2017} and utilized for controlling transition to turbulence \citep{Kawahara2005}.
The energy of the edge state, together with that of \textit{minimal seeds}, i.e., that of the minimal energy perturbations that trigger turbulence \citep{Pringle2012, Duguet2013}, can therefore be thought of as main scalar indicators of the robustness of the laminar flow to finite amplitude perturbations.

A tempting way to design control strategies for transition is via the maximization of the energy of the minimal seed \citep{Rabin2014} or of the edge state.
In this paper, we show that such quantities may be insufficient to provide a reliable conclusion about the efficiency of control strategies.
To address such shortcomings, we introduce a novel methodology: the statistical sampling of the phase space to determine the probability that perturbations laminarize as a function of their energy.
We apply this approach in plane Couette flow where control is imposed via transverse wall oscillations \citep{Jung1992, Baron1995, Quadrio2004, Rabin2014}.

\section{Protocol}
We consider plane Couette flow, i.e., the flow confined between two infinite plates separated by a gap $2h$ and moving in opposite directions with constant velocity $U$.
The domain is periodic in the streamwise (period $\Gamma_x = 4 \pi h$) and in the spanwise (period $\Gamma_z = 32 \pi h /15$) directions, and no-slip boundary conditions are applied at the walls \citep{Pershin2019}.
The nondimensionalized Navier--Stokes equation and the incompressibility condition are given by:
\begin{eqnarray}
  \label{navsto}
&\partial_t \boldsymbol{u} + v \boldsymbol{e_x} + y \partial_x \boldsymbol{u} + (\boldsymbol{u} \cdot \nabla) \boldsymbol{u} = -\nabla p + \displaystyle\frac{1}{Re} \nabla^2 \boldsymbol{u},\\
  \label{incomp}
&\nabla \cdot \boldsymbol{u} = 0,
\end{eqnarray}
where $Re = U h/\nu$ is the Reynolds number, $\nu$ is the kinematic viscosity, $p$ is the pressure, $\boldsymbol{e_x}$ is the unit vector in the streamwise direction and $\boldsymbol{u}$ satisfies homogeneous boundary conditions in $y$.
To obtain these equations, we have decomposed the nondimensional velocity $\boldsymbol{U} = \boldsymbol{U_{lam}} + \boldsymbol{u}$, where the laminar flow $\boldsymbol{U_{lam}} = y \boldsymbol{e_x}$ and $\boldsymbol{u}$ is the incompressible perturbation.
The no-slip boundary conditions read: $\boldsymbol{u} (y=\pm 1) = \boldsymbol{0}$.

We use \textit{Channelflow} \citep{Gibson2014channelflow} to solve for the flow at various values of the Reynolds number $Re > Re_c$, where $Re_c = 325 \pm 10$ is the critical value above which sustained turbulence can be observed \citep{Shi2013, Dauchot1995finite}, and in the presence of control by wall oscillations at $Re = 500$.
The streamwise and spanwise coordinates are discretized using $N_x=32$ and $N_z=34$ Fourier coefficients and the wall-normal coordinate is discretized using $N_y=33$ Chebyshev coefficients based on the numerical resolution used in \cite{Pershin2019}.
The temporal discretization is performed by means of 3rd-order semi-implicit backward differentiation with time step $\triangle t = 1 / \Rey$.
For $Re = 500$, the time-averaged kinetic energy of turbulent flow realizations is $E_{turb} \approx 6 \times 10^{-2}$, where the instantaneous kinetic energy $E$ for the state $\boldsymbol{u}$ is defined as
\begin{equation}
\label{eq:E_def}
E = \frac{1}{2}\langle \boldsymbol{u}, \boldsymbol{u} \rangle = \frac{1}{2}||\boldsymbol{u}||^2 = \frac{1}{2V}\int_{\Omega} \boldsymbol{u} \cdot \boldsymbol{u} \; \partial \Omega,
\end{equation}
where $V = 2 \Gamma_x \Gamma_z = 256 \pi^2/15$ is the volume of the domain $\Omega$.
We time-integrate a number of initial conditions and assess their behavior in the following way.
If the kinetic energy of the resulting flow decays below $E_{lam} = E_{turb} / 100$ for at most $t_{turb} = 400$ time units, the flow is said to have laminarized.
Conversely, if the energy exceeds $E_{turb}$, the flow is said to have transitioned.
The role of the non-zero waiting time $t_{turb}$ is to ensure that an event where the flow exceeds $E_{turb}$ and decays immediately after is not counted as transitional.
Such events associated with high-energy perturbations may be due to the crossing of the \textit{upper edge of chaos} \citep{Budanur2020}.

We compute $P_{lam}(E^{(j)})$, the laminarization probability of a random initial perturbation (RP) of energy $E^{(j)}$.
To do so, we consider $40$ energy levels $E^{(j)}, j = 1, \dots, 40$, equispaced between $0$ and $(2/3) E_{turb}$.
\begin{figure*}
    \includegraphics[width=1.0\textwidth]{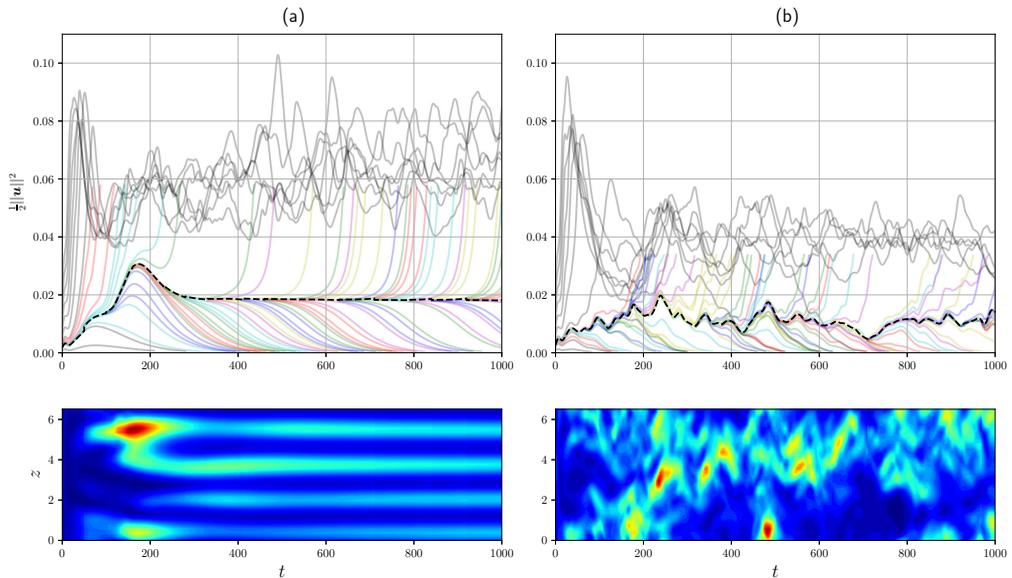}
    \caption{Edge tracking results for (a) the uncontrolled and (b) the controlled cases at $Re = 500$ shown via the evolution of the norm $||\boldsymbol{u}||^2/2$ in time (top panels). A set of initial conditions differing only by their amplitude are integrated forward in time (evolution shown in gray lines) to find the two closest trajectories that display different behavior. At their point of divergence, both flow states are fed into the bisection-type algorithm in a repeated process \citep{Skufca2006}. Successive iterations are shown in light colors, together with the resulting edge trajectory in thick dashed line. The time-evolution of the kinetic energy of the edge trajectories, averaged in the streamwise and wall-normal directions, is shown on the bottom panels for (a) the uncontrolled and (b) the controlled cases. Each edge trajectory approaches the corresponding edge state which appears to be an equilibrium in the uncontrolled case (a) and a chaotic object in the controlled case (b).}
    \label{fig:edgetracking}
\end{figure*}
For each energy level, we generate $N$ RPs, $\boldsymbol{u}^{(j, k)}, k = 1, \dots, N$, which we time-integrate until one of the aforementioned energy thresholds is crossed to determine the type of flow.
We use $N = 100$ RPs per energy level, except for the $Re = 500$ case for which we considered $N = 200$ to generate more accuracy and allow for a better assessment of the control strategy.
We approximate $P_{lam}(E^{(j)})$ by the fraction of laminarizing RPs of energy $E^{(j)}$.
One may observe that due to the finite lifetime of turbulent trajectories in shear flows in small domains \citep{Hof2006}, the laminarization probability may depend on the choice of $t_{turb}$.
The transient lifetimes that we observed were all, at least, an order of magnitude larger than $t_{turb}$, making our results independent on this value.
We also computed the edge state kinetic energy $E_{edge}$ at $Re = 400, 500$ and $700$ using \textit{edge tracking} \citep{Skufca2006} and report the corresponding values in table \ref{tb:summary}.
We demonstrate the edge tracking procedure at $Re = 500$ in figure \ref{fig:edgetracking}(a) which yields $E_{edge} \approx 1.82 \times 10^{-2} \approx (1/3) E_{turb}$.
As can be seen from the the bottom plane in figure \ref{fig:edgetracking}(a), the edge state for the uncontrolled case is an equilibrium structurally similar to the edge state identified in \cite{Schneider2008} in a similar domain.

To generate RPs, we first express them as a linear combination of the laminar flow field $\boldsymbol{U_{lam}}$ and an incompressible orthogonal component $\boldsymbol{u_{\bot}}^{(j, k)}$, i.e.\ $\boldsymbol{u}^{(j, k)} = A \boldsymbol{u_{\bot}}^{(j, k)} + B \boldsymbol{U_{lam}}$, where $A$, $B$ and $\boldsymbol{u_{\bot}}^{(j, k)}$ are generated randomly, $\langle \boldsymbol{u_{\bot}}^{(j, k)},\boldsymbol{U_{lam}}\rangle = 0$ and $||\boldsymbol{u_{\bot}}^{(j, k)}||~=~1$.
To ensure that $\boldsymbol{u}^{(j, k)}$ has energy $E^{(j)}$, we further impose $||A \boldsymbol{u_{\bot}}^{(j, k)} + B \boldsymbol{U_{lam}}||^2 = 2 E^{(j)}$.
Turbulence in plane Couette flow is associated with shear concentration at the wall, yielding lower kinetic energies for the turbulent states than for the laminar flow.
We thus find it intuitively useful to distinguish RPs with weakened bulk shear ($B<0$) from those that display stronger bulk shear ($B \ge 0$), the sets of which are hereafter called RP$-$ and RP$+$ respectively.
The former can be thought of as being mostly located in phase space between the laminar (energetically higher) and the turbulent (energetically lower) flows, while the latter are located farther away from turbulence.
We create such perturbations in three steps.
First, we generate the random orthogonal component $\boldsymbol{u_{\bot}}^{(j, k)}$ by drawing its spectral coefficients from the uniform distribution so that the homogeneous boundary conditions are satisfied and $||\boldsymbol{u_{\bot}}^{(j, k)}|| = 1$.
Next, we draw $B$ from the uniform distribution between $- 2 E^{(j)} / ||\boldsymbol{U_{lam}}||$ and $2 E^{(j)} / ||\boldsymbol{U_{lam}}||$ and compute $A = \pm \sqrt{2 E^{(j)} - B^2 ||\boldsymbol{U_{lam}}||^2}$.
To ensure that the RPs satisfy the no-slip boundary condition, we take advantage of the properties of our numerical scheme (Spalart--Moser--Rogers Runge--Kutta, see \cite{Spalart91}; leading to spatial operators solved using a Chebyshev tau algorithm, see \cite{chqz88}) and take two (small) time-steps starting from $\boldsymbol{u}^{(j, k)}$. The resulting state is RP $\boldsymbol{u}^{(j, k)}$ \footnote{For small enough time-steps, this procedure leads to negligible deviations in the energy level and in the coefficients $A$ and $B$ of the RP.}.
To ensure fair sampling, we use an equal number of positive and negative $B$ for each energy level.

\section{Results}
\begin{figure}
    \centering
    \includegraphics[width=1.0\textwidth]{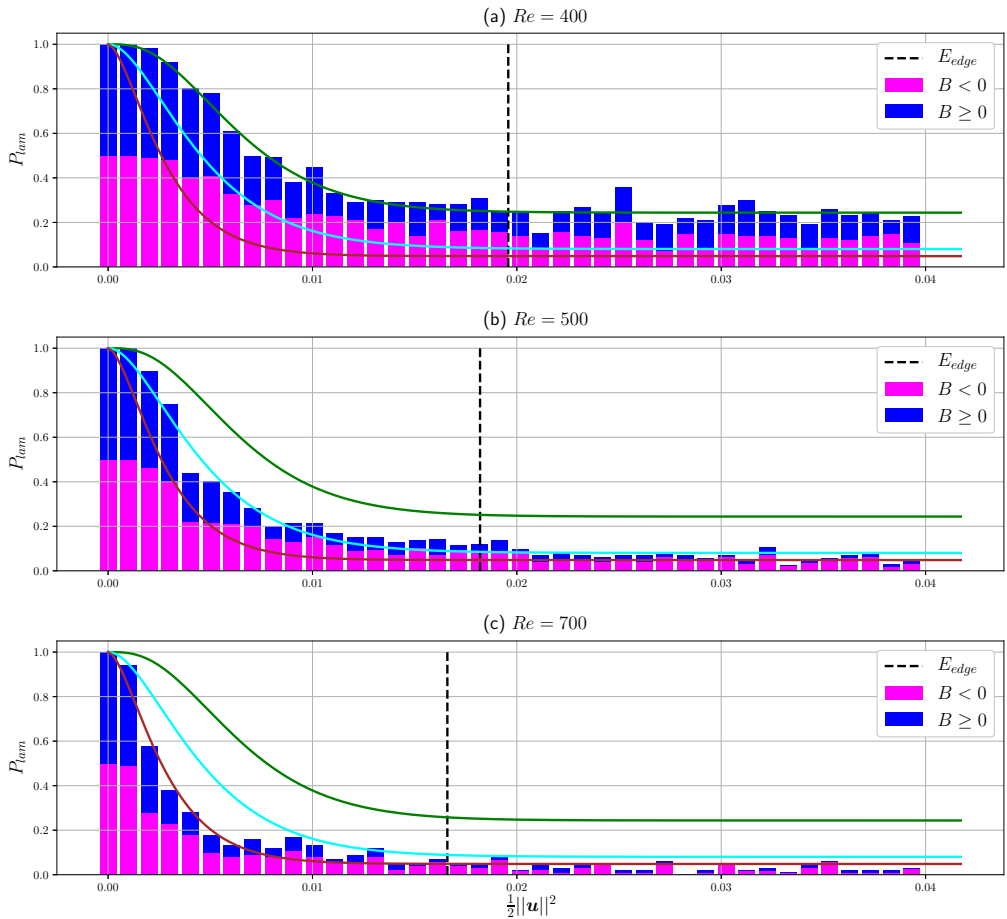}
    \caption{Laminarization probability $P_{lam}$ for $40$ equispaced energy levels from $0$ to $0.04$ in plane Couette flow calculated for three Reynolds numbers: $Re = 400$ (a), $Re = 500$ (b) and $Re = 700$ (c). The probability contributions from RP$-$ (resp. RP$+$) are shown in magenta (resp. blue). The solid fitting curves (green for $Re = 400$, cyan for $Re = 500$ and brown for $Re = 700$) correspond to $p(E)$ explained in the text. The vertical dashed lines indicate the energy of the edge state at a particular $Re$.}
    \label{fig:p_lam}
\end{figure}

\begin{figure}
    \centering
    \includegraphics[width=1.0\textwidth]{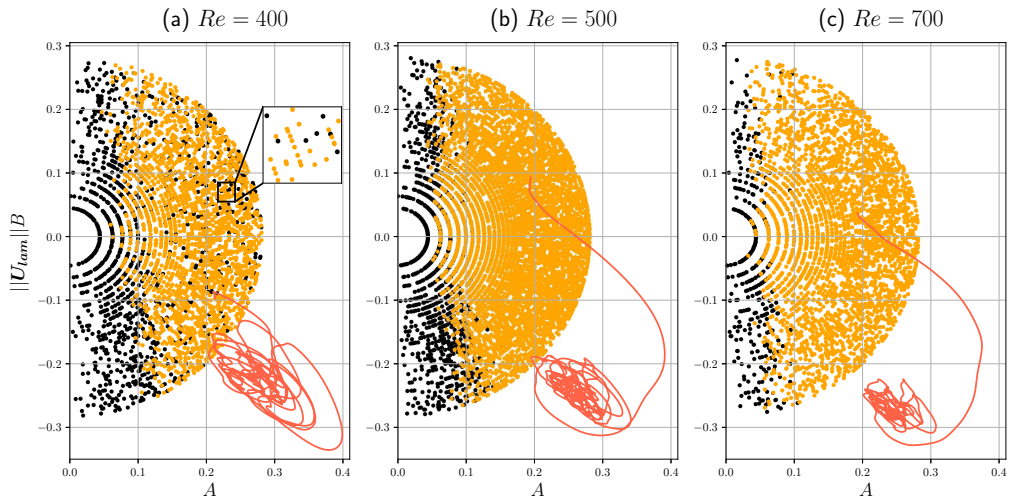}
    \caption{Representation of all the initial perturbations considered in figure \ref{fig:p_lam} in the $A$-$||\boldsymbol{U_{lam}}||B$ plane shown for $Re = 400$ (a), $Re = 500$ (b) and $Re = 700$ (c). Laminarizing (resp. transitioning) perturbations are shown in black (resp. orange). Energy levels correspond to half-circles centered at $A=B=0$. Examples of turbulent trajectories emanating from transitioning perturbations and time-integrated for $t = 2000$ time units are shown by red curves for each considered Reynolds number. Note that the distribution of the points in the middle plot is denser since we used $N = 200$ RPs per energy level at $Re = 500$.}
    \label{fig:ics}
\end{figure}

The results in the absence of any transition control are reported in figure \ref{fig:p_lam} via the laminarization probability as a function of the RP energy at $Re = 400, 500$ and $700$.
We start the description from the $Re = 500$ case.
For sufficiently small initial disturbance energies, the laminarization probability tends to 1, owing to the fact that the laminar flow is linearly stable.
The first transitioning RP was found at the third energy level $E^{(3)} \approx 4 \times 10^{-3}$ providing an upper bound for the minimal seed energy, $E_{min} \le 4 \times 10^{-3}$, so that the minimal seed is located very close to the laminar fixed point.
The laminarization probability decreases nearly monotonically as the RP energy increases to saturate at $P_{lam} \approx 0.08$ which we will hereafter quantify by $a$.
Most of the laminarization probability decay occurs at small amplitude, i.e., for $E < E_{edge}$ ($P_{lam} (E_{edge}) \approx 0.12$).
For the most part, RP$-$ is responsible for the non-vanishing laminarization probability as the energy increases.
The fact that $P_{lam}$ does not tend to zero even for large energies is a consequence of the edge structure: it has been shown to be wrapped around the turbulent saddle such that laminarizing RP regions are locally interleaved with transitioning RP regions \citep{Chantry2014}, and to be fractal \citep{Moehlis2004, Skufca2006}.
To understand further which RPs laminarize, we color-code them in the $A$-$||\boldsymbol{U_{lam}}||B$ plane according to their dynamics.
The results for $Re = 500$ are shown in figure \ref{fig:ics}(b).

Laminarizing RPs are found for small $A$ and their number decreases as $B$ increases.
Outside this clearly identifiable region, we observe very rare transitional events, evidencing the usefulness of our decomposition of the initial condition for the study of the robustness of the laminar flow.
In addition, figure \ref{fig:ics}(b) shows a typical trajectory undertaken by a transitioning perturbation.
This turbulent trajectory gets trapped in a small region located in the lower right corner of the figure and associated with the projection of the turbulent saddle onto the reduced subspace.
This region corresponds to negative $B$ which confirms the physical reasoning behind our decomposition.

We now explore the dependence of these results on the Reynolds number.
Increasing the Reynolds number from $500$ to $700$ (figures \ref{fig:p_lam}(c) and \ref{fig:ics}(c)) reduces the upper bound for the minimal seed energy which reflects the fact that the minimal seed energy exhibits a power-law decay as a function of $Re$ \citep{Duguet2013}.
Similarly, increasing $Re$ reduces the edge state energy, in agreement with past studies \citep{Wang2007}.
The probability associated with the distribution plateau also decreases as the Reynolds number increases ($a = 0.244$ at $Re = 400$ and $a = 0.0484$ at $Re = 700$) implying that less large-energy RPs laminarize.
In general, the distribution $P_{lam}(E)$ becomes more peaked at $E = 0$ as $Re$ increases which reflects the anticipated shrinkage of the basin of attraction of the laminar flow as $Re$ is increased away from criticality.
We can analyse the evolution of the basin of attraction of the laminar flow with $Re$ by inspecting the distribution of laminarizing and transitioning RPs in the $A$-$||\boldsymbol{U_{lam}}||B$ plane in figure \ref{fig:ics}.
At $Re = 700$, the laminarizing RPs are heavily concentrated in the region around $A = 0$.
Decreasing $Re$ slightly expands this region and, close to criticality, leads to the appearance of laminarizing RPs outside this region.
Additionally, the representative transitioning trajectories shown in figure \ref{fig:ics} indicate that the turbulent saddle expands in our projection of phase space as the Reynolds number is decreased.
This reflects the fact that the variance of the turbulent kinetic energy grows as the Reynolds number is reduced towards the criticality \citep{Faranda2014}.
Finally, figures \ref{fig:p_lam} and \ref{fig:ics} suggest that RP$+$ behaves similarly to RP$-$ for $Re=400$ but that their laminarization probability tends to decrease much faster than that of RP$-$ as the Reynolds number is increased.

The probability distribution for the uncontrolled case can be approximated by the cumulative distribution function for the Gamma distribution reflected around $0.5$ and saturated at $a$, i.e. $p(E) = 1 - (1 - a)\gamma(\alpha, \beta E)$, where $\gamma(\alpha, \beta E)$ is the lower incomplete gamma function and the values of $\alpha$, $\beta$ and $a$ are reported in table \ref{tb:summary}.
The various coefficients have been determined via least-square fitting and the resulting functions are shown in figure \ref{fig:p_lam} by the solid lines.
One can assess control strategies by simple quantitative comparison with the distribution at a particular Reynolds number.
If the action of a control strategy leads to an increase of the laminarization probability, it is successful.
The difference in the shape of the laminarization probability reveals important information about the sensitivity of the laminar flow to perturbations of various amplitudes and can be used as a measure for the relative increase of the size of the basin of attraction of the laminar flow thereby quantifying control efficiency.

\begin{table}
\begin{center}
\def~{\hphantom{0}}
\begin{tabular}{lccc|c}
 & $Re = 400$ & $Re = 500$ & $Re = 700$ & $Re = 500$, controlled \\
$a$ & $0.244$ & $0.0805$ & $0.0484$ & $0.286$ \\
$\alpha$ & $3.43$ & $2.05$ & $1.79$ & $3.75$ \\
$\beta$ & $500$ & $412$ & $593$ & $899$ \\
$E_{edge}$ & $1.96 \times 10^{-2}$ & $1.82 \times 10^{-2}$ & $1.66 \times 10^{-2}$ & $1.15 \times 10^{-2}$
\end{tabular}
\caption{Values of parameters $a, \alpha$ and $\beta$ of the fitting function $p(E)$, as explained in the text, and of the time-averaged edge state energy $E_{edge}$. These values are reported in $4$ cases: for the uncontrolled system at $Re = 400, 500$ and $700$ and for the controlled system (spanwise wall oscillations with amplitude $W = 0.3$ and frequency $\omega = 1/16$) at $Re = 500$.}
\label{tb:summary}
\end{center}
\end{table}

To demonstrate this, we impose in-phase spanwise wall oscillations, a strategy known to reduce the turbulent drag \citep{Quadrio2004} and increase the energy of the minimal seed \citep{Rabin2014}.
Under these oscillations, the modified boundary conditions read: $\boldsymbol{U}(y = \pm 1) = [\pm 1, 0, W \sin(\omega t + \phi)]$, where $W$, $\omega$ and $\phi \in [0; 2\pi)$ are the amplitude, the frequency and the phase of the oscillations.
The laminar flow becomes oscillatory, which modifies equation (\ref{navsto}) into equation (2.5) of \cite{Rabin2014}.
We use the parameter values close to the ones considered by \cite{Rabin2014}, $W = 0.3$ and $\omega = 1/16$, and set $Re = 500$.
For these parameter values and domain size, we find that the edge state is chaotic, as shown in figure \ref{fig:edgetracking}(b), with average kinetic energy $E_{edge}^{(osc)} \approx 1.15 \times 10^{-2}$, approximately $37\%$ less than for the uncontroled edge state.
We generate the RPs in the same way as for the uncontrolled case, except that we also impose a random phase $\phi$ drawn from a uniform distribution between $0$ and $2 \pi$.

\begin{figure}
    \centering
    \includegraphics[width=1.0\textwidth]{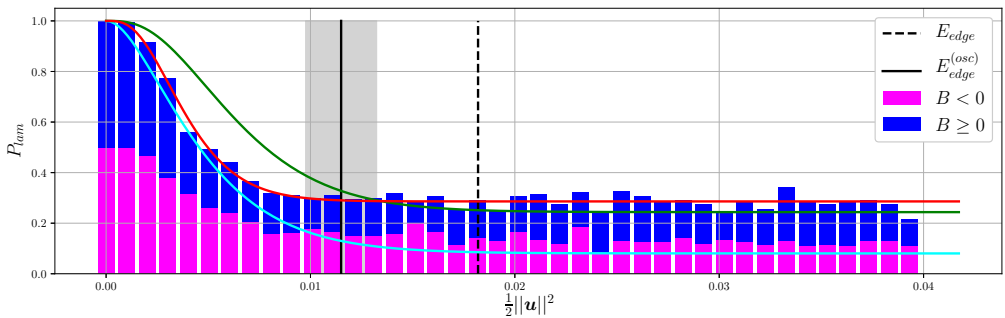}
    \caption{Same as figure \ref{fig:p_lam}(b) but with the oscillating wall control as explained in the text. The green and cyan curves correspond to the fitting curves obtained in figure \ref{fig:p_lam} for the uncontrolled case at $Re = 400$ and $Re = 500$ respectively, whereas the red fitting curve corresponds to that of the controlled case, $p_{osc}(E)$, as explained in the text. The region of standard deviation of the edge state energy about its mean value is shaded in gray.}
    \label{fig:p_lam_osc}
\end{figure}

The laminarization probability for the controlled case is shown in figure \ref{fig:p_lam_osc}.
The resulting probability distribution decreases nearly monotonically to saturate around $P_{lam} = 0.3$ for large energy RPs.
This behavior is qualitatively similar to that observed in the absence of control but fundamental quantitative differences can be reported.
Firstly, the probability distribution plateaus at lower RP energy than in the uncontrolled case.
This results in a larger asymptotic value of $P_{lam}$, more than double its value in the absence of wall oscillation.
Secondly, a non-negligible fraction of large energy initial perturbations of RP$+$ are now found to laminarize.

that of reducing the Reynolds number in the absence of control (see figure 2(a) for Re = 400)

\begin{figure}
    \centering
    \includegraphics[width=0.5\textwidth]{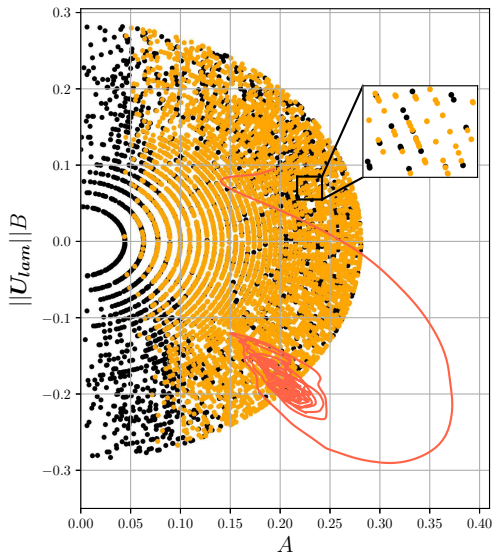}
    \caption{Same as figure \ref{fig:ics}(b) but with the oscillating wall control as explained in the text.}
    \label{fig:ics_osc}
\end{figure}

The effect of applying this control strategy onto transition at $Re= 500$ seems to be similar to that of reducing the Reynolds number in the absence of control (see figure \ref{fig:p_lam}(a) for $Re = 400$) despite the fact that wall oscillations effectively increase the Reynolds number by making the walls move in the spanwise direction in addition to their steady motion in the streamwise direction.
An important difference can however be highlighted: this control strategy did not statistically affect the behavior of small energy perturbations from the laminar flow.
To shed more light on the effect of control on the distribution of the laminarizing and transitioning RPs, we show, in figure \ref{fig:ics_osc}, a similar representation of the initial conditions as in figure \ref{fig:ics}(b) but for the controlled case.
We recover the small $A$ region where laminarization was found in the uncontrolled case, however, we also found laminarizing RPs at larger amplitude, scattered around the region where all RPs transitioned in the absence of control.
This result also bears resemblance with that at $Re = 400$ in the absence of control (see figure \ref{fig:ics}(a)).
The newly controlled RPs decay via overshooting, i.e., their energy first significantly exceeds $E_{turb}$ before decaying nearly monotonically.
The initial phase $\phi$ of the RPs did not seem to play any role in determining whether or not the flow will laminarize.

The laminarization probability in the controlled case can be approximated by the fitting function $p_{osc}(E)$, which shares the same structure as $p(E)$ but with parameter values shown in table \ref{tb:summary}.
The relative probability increase can thus be computed as $(p_{osc}(E) - p(E)) / p(E)$ averaged over the range of the considered energies.
We found that this quantity is equal to $1.8$ so that, on average, the laminarization probability of an RP nearly doubles under the action of the aforementioned control strategy.
When a single value is not satisfactory to assess the control efficiency, more detailed information can be obtained via inspecting the differences between the laminarization probabilities at each energy level.

While the fate of perturbations in the uncontrolled flow at $Re = 500$ is sensitive to the strength of the initial bulk shear measured by parameter $B$, this is no longer the case in the presence of control via spanwise wall oscillation.
Furthermore, the energy of the edge state decreases under the effect of the oscillating wall.
Assuming that the edge state energy is proportional to the relative volume of the basin of attraction of the laminar fixed point, this could be interpreted as the failure of the control strategy to postpone transition but the more exhaustive analysis of phase space provided here shows that this strategy is on the contrary effective.
The use of the minimal seed does not seem to provide a better basis for control assessment: the shape of the basin of attraction of the laminar flow is such that, in this study, most perturbations generated with $4$ times the energy of the minimal seed laminarized, making it difficult to extrapolate any reliable information.
These observations further strengthen our approach to consider a more exhaustive method to assess control.

\section{Discussion}
In this paper, we have introduced a new way to analyze the robustness of the laminar flow to perturbations and to assess control strategies.
We proceeded by sampling phase space in the neighbourhood of the laminar flow and evaluating the probability that the sampled initial conditions laminarize as a function of their initial energy.
Our laminarisation probability bears similarities with the notion of basin stability introduced in the dynamical system context of vegetation growth \citep{Menck2013}.
Our results for plane Couette flow in a small domain indicate that the laminarization probability decreases as the kinetic energy of the initial perturbation from the laminar flow is increased.
It may increase back to $1$ for sufficiently large initial energy levels as a consequence of the location of the upper edge of chaos \citep{Budanur2020}.
We identified that the laminarization probability decreases as the Reynods number is increased which reflects the contraction of the basin of attraction of the laminar flow.
Two plausible scalar proxies for the relative volume of this basin, the edge state and the minimal seed energies, though having different asymptotic scalings with respect to $Re$, similarly reflect this expansion: their values increase as $Re$ is reduced towards the criticality \citep{Wang2007, Duguet2013}.

To assess how the basin of attraction of the laminar flow changes under the action of control, the distribution of the laminarization probability obtained in the uncontrolled set-up can easily be recomputed and compared in the presence of control.
We tested this methodology under control via spanwise wall oscillations at $Re = 500$.
It was shown that, under the action of similar control, the minimal seed energy increases \cite{Rabin2014} while we observed that the energy of the edge state decreases.
The use of the newly introduced laminarization probability provides a much more in-depth understanding of the alteration of the robustness of the laminar flow.
In particular, it revealed that the control strategy under consideration provided a major improvement in the robustness of the laminar flow, especially against large-energy perturbations.
The controlled case gave similar results to the uncontrolled flow at $Re = 400$.
The laminarization probabilities differ in a subtle way: the control strategy is not efficient in laminarizing small-energy RPs but acts favourably on large-energy RPs.
These observations suggest that, in contrast to the laminarization probability, scalar criteria for the robustness of the laminar flow, such as the energy of the edge state or of the minimal seed, may not be able to capture enough information to assess control strategies -- the fact already noted for the edge states in the polymer drag reduction studies \citep{Stone2002, Stone2004}.

In setting up our protocol, we had to make a choice on the form of the random initial perturbation.
We opted for randomizing their spectral coefficients uniformly, which might not be the ideal choice if one knows what form of perturbations are triggered in a given configuration.
We stress that the current work represents a proof of concept to pave the way to customizable and more relevant protocols.
There are many possibilities to constrain the initial perturbations and increase their relevance to given situations.
For example, one can impose a certain shape for the spectral energy of the initial perturbations, or confine them to certain locations in the physical domain.

We believe that the methodology introduced in this paper will prove useful to control other flows \citep{Khapko2013,Zammert2015,Watanabe2016,Chantry2017} and anticipate that, owing to the nature of transition to turbulence, it will be helpful to the study and control of a range of nonlinear systems displaying finite-amplitude instability.

\section*{Acknowledgments}

This work was undertaken on ARC3, part of the High Performance Computing facilities at the University of Leeds, UK.
The authors wish to thank C. Bick and R. Kerswell for fruitful discussions and the anonymous referees for their suggestions which led to a significant improvement of the paper.

\section*{Declaration of Interests}
The authors report no conflict of interest.

\bibliographystyle{jfm}
\bibliography{2019_JFM_A_probabilistic_protocol}

\begin{thebibliography}{41}
\expandafter\ifx\csname natexlab\endcsname\relax\def\natexlab#1{#1}\fi
\def\au#1{#1} \def\ed#1{#1} \def\yr#1{#1}\def\at#1{#1}\def\jt#1{\textit{#1}}
  \def\bt#1{#1}\def\bvol#1{\textbf{#1}} \def\vol#1{#1} \def\pg#1{#1}
  \def\publ#1{#1}\def\arxiv#1{#1}\def\org#1{#1}\def\st#1{\textit{#1}}

\bibitem[Avila {\em et~al.\/}(2011)Avila, Moxey, de~Lozar, Avila, Barkley \&
  Hof]{Avila2011}
{\sc \au{Avila, K.}, \au{Moxey, D.}, \au{de~Lozar, A.}, \au{Avila, M.},
  \au{Barkley, D.} \& \au{Hof, B.}} \yr{2011}  \at{The onset of turbulence in
  pipe flow}.  \jt{Science}  \bvol{333}~(6039),  \pg{192--196}.

\bibitem[Baron \& Quadrio(1995)]{Baron1995}
{\sc \au{Baron, A.} \& \au{Quadrio, M.}} \yr{1995}  \at{Turbulent drag
  reduction by spanwise wall oscillations}.  \jt{Appl. Sci. Res.}
  \bvol{55}~(4),  \pg{311--326}.

\bibitem[Brunton \& Noack(2015)]{Brunton2015}
{\sc \au{Brunton, S.~L.} \& \au{Noack, B.~R.}} \yr{2015}  \at{Closed-loop
  turbulence control: progress and challenges}.  \jt{Appl. Mech. Rev.}
  \bvol{67}~(5).

\bibitem[Budanur {\em et~al.\/}(2020)Budanur, Marensi, Willis \&
  Hof]{Budanur2020}
{\sc \au{Budanur, N.~B.}, \au{Marensi, E.}, \au{Willis, A.~P.} \& \au{Hof, B.}}
  \yr{2020}  \at{Upper edge of chaos and the energetics of transition in pipe
  flow}.  \jt{Phys. Rev. Fluids}  \bvol{5}~(2),  \pg{023903}.

\bibitem[Canuto {\em et~al.\/}(1988)Canuto, Hussaini, Quarteroni \&
  Zang]{chqz88}
{\sc \au{Canuto, C.}, \au{Hussaini, M.~Y.}, \au{Quarteroni, A.} \& \au{Zang,
  T.~A.}} \yr{1988}  \at{Spectral methods -- {F}undamentals in single domains}.
   \jt{Springer-Verlag, New York} .

\bibitem[Chantry \& Schneider(2014)]{Chantry2014}
{\sc \au{Chantry, M.} \& \au{Schneider, T.~M.}} \yr{2014}  \at{Studying edge
  geometry in transiently turbulent shear flows}.  \jt{J. Fluid Mech.}
  \bvol{747},  \pg{506--517}.

\bibitem[Chantry {\em et~al.\/}(2017)Chantry, Tuckerman \&
  Barkley]{Chantry2017}
{\sc \au{Chantry, M.}, \au{Tuckerman, L.~S.} \& \au{Barkley, D.}} \yr{2017}
  \at{Universal continuous transition to turbulence in a planar shear flow}.
  \jt{J. Fluid Mech.}  \bvol{824},  \pg{R1}.

\bibitem[Dauchot \& Daviaud(1995)]{Dauchot1995finite}
{\sc \au{Dauchot, O.} \& \au{Daviaud, F.}} \yr{1995}  \at{Finite amplitude
  perturbation and spots growth mechanism in plane {C}ouette flow}.  \jt{Phys.
  Fluids}  \bvol{7}~(2),  \pg{335--343}.

\bibitem[Duguet {\em et~al.\/}(2010)Duguet, Brandt \& Larsson]{Duguet2010}
{\sc \au{Duguet, Y.}, \au{Brandt, L.} \& \au{Larsson, B. R.~J.}} \yr{2010}
  \at{Towards minimal perturbations in transitional plane {C}ouette flow}.
  \jt{Phys. Rev. E}  \bvol{82}~(2),  \pg{026316}.

\bibitem[Duguet {\em et~al.\/}(2013)Duguet, Monokrousos, Brandt \&
  Henningson]{Duguet2013}
{\sc \au{Duguet, Y.}, \au{Monokrousos, A.}, \au{Brandt, L.} \& \au{Henningson,
  D.~S.}} \yr{2013}  \at{Minimal transition thresholds in plane {C}ouette
  flow}.  \jt{Phys. Fluids}  \bvol{25}~(8),  \pg{084103}.

\bibitem[Faranda {\em et~al.\/}(2014)Faranda, Lucarini, Manneville \&
  Wouters]{Faranda2014}
{\sc \au{Faranda, D.}, \au{Lucarini, V.}, \au{Manneville, P.} \& \au{Wouters,
  J.}} \yr{2014}  \at{On using extreme values to detect global stability
  thresholds in multi-stable systems: The case of transitional plane {C}ouette
  flow}.  \jt{Chaos, Solitons \& Fractals}  \bvol{64},  \pg{26--35}.

\bibitem[Gibson(2014)]{Gibson2014channelflow}
{\sc \au{Gibson, J.~F.}} \yr{2014}  \bt{{Channelflow}: {A} spectral
  {Navier--Stokes} simulator in {C}++}. {\em Tech. Rep.\/}.  \org{U. New
  Hampshire}, {\tt {Channelflow.org}}.

\bibitem[Hof {\em et~al.\/}(2010)Hof, De~Lozar, Avila, Tu \&
  Schneider]{Hof2010}
{\sc \au{Hof, B.}, \au{De~Lozar, A.}, \au{Avila, M.}, \au{Tu, X.} \&
  \au{Schneider, T.~M.}} \yr{2010}  \at{Eliminating turbulence in spatially
  intermittent flows}.  \jt{Science}  \bvol{327}~(5972),  \pg{1491--1494}.

\bibitem[Hof {\em et~al.\/}(2006)Hof, Westerweel, Schneider \&
  Eckhardt]{Hof2006}
{\sc \au{Hof, B.}, \au{Westerweel, J.}, \au{Schneider, T.~M.} \& \au{Eckhardt,
  B.}} \yr{2006}  \at{Finite lifetime of turbulence in shear flows}.
  \jt{Nature}  \bvol{443}~(7107),  \pg{59}.

\bibitem[Jung {\em et~al.\/}(1992)Jung, Mangiavacchi \& Akhavan]{Jung1992}
{\sc \au{Jung, W.-J.}, \au{Mangiavacchi, N.} \& \au{Akhavan, R.}} \yr{1992}
  \at{Suppression of turbulence in wall-bounded flows by high-frequency
  spanwise oscillations}.  \jt{Phys. Fluids A-Fluid}  \bvol{4}~(8),
  \pg{1605--1607}.

\bibitem[Kasagi {\em et~al.\/}(2009)Kasagi, Suzuki \& Fukagata]{Kasagi2009}
{\sc \au{Kasagi, N.}, \au{Suzuki, Y.} \& \au{Fukagata, K.}} \yr{2009}
  \at{Microelectromechanical systems--based feedback control of turbulence for
  skin friction reduction}.  \jt{Annu. Rev. Fluid Mech.}  \bvol{41}.

\bibitem[Kawahara(2005)]{Kawahara2005}
{\sc \au{Kawahara, G.}} \yr{2005}  \at{Laminarization of minimal plane
  {C}ouette flow: going beyond the basin of attraction of turbulence}.
  \jt{Phys. Fluids}  \bvol{17}~(4),  \pg{041702}.

\bibitem[Khapko {\em et~al.\/}(2013)Khapko, Kreilos, Schlatter, Duguet,
  Eckhardt \& Henningson]{Khapko2013}
{\sc \au{Khapko, T.}, \au{Kreilos, T.}, \au{Schlatter, P.}, \au{Duguet, Y.},
  \au{Eckhardt, B.} \& \au{Henningson, D.~S.}} \yr{2013}  \at{Localized edge
  states in the asymptotic suction boundary layer}.  \jt{J. Fluid Mech.}
  \bvol{717},  \pg{R6}.

\bibitem[Khapko {\em et~al.\/}(2016)Khapko, Kreilos, Schlatter, Duguet,
  Eckhardt \& Henningson]{Khapko2016}
{\sc \au{Khapko, T.}, \au{Kreilos, T.}, \au{Schlatter, P.}, \au{Duguet, Y.},
  \au{Eckhardt, B.} \& \au{Henningson, D.~S.}} \yr{2016}  \at{Edge states as
  mediators of bypass transition in boundary-layer flows}.  \jt{J. Fluid Mech.}
   \bvol{801}.

\bibitem[Kim \& Bewley(2007)]{Kim2007}
{\sc \au{Kim, J.} \& \au{Bewley, T.~R.}} \yr{2007}  \at{A linear systems
  approach to flow control}.  \jt{Annu. Rev. Fluid Mech.}  \bvol{39},
  \pg{383--417}.

\bibitem[Marensi {\em et~al.\/}(2019)Marensi, Willis \& Kerswell]{Marensi2019}
{\sc \au{Marensi, E.}, \au{Willis, A.~P.} \& \au{Kerswell, R.~R.}} \yr{2019}
  \at{Stabilisation and drag reduction of pipe flows by flattening the base
  profile}.  \jt{J. Fluid Mech.}  \bvol{863},  \pg{850--875}.

\bibitem[Mellibovsky {\em et~al.\/}(2009)Mellibovsky, Meseguer, Schneider \&
  Eckhardt]{Mellibovsky2009}
{\sc \au{Mellibovsky, F.}, \au{Meseguer, A.}, \au{Schneider, T.~M.} \&
  \au{Eckhardt, B.}} \yr{2009}  \at{Transition in localized pipe flow
  turbulence}.  \jt{Phys. Rev. Lett.}  \bvol{103}~(5),  \pg{054502}.

\bibitem[Menck {\em et~al.\/}(2013)Menck, Heitzig, Marwan \& Kurths]{Menck2013}
{\sc \au{Menck, P.~J.}, \au{Heitzig, J.}, \au{Marwan, N.} \& \au{Kurths, J.}}
  \yr{2013}  \at{How basin stability complements the linear-stability
  paradigm}.  \jt{Nat. Phys.}  \bvol{9}~(2),  \pg{89--92}.

\bibitem[Moehlis {\em et~al.\/}(2004)Moehlis, Faisst \& Eckhardt]{Moehlis2004}
{\sc \au{Moehlis, J.}, \au{Faisst, H.} \& \au{Eckhardt, B.}} \yr{2004}  \at{A
  low-dimensional model for turbulent shear flows}.  \jt{New J. Phys.}
  \bvol{6}~(1),  \pg{56}.

\bibitem[Olvera \& Kerswell(2017)]{Olvera2017}
{\sc \au{Olvera, D.} \& \au{Kerswell, R.~R.}} \yr{2017}  \at{Optimizing energy
  growth as a tool for finding exact coherent structures}.  \jt{Phys. Rev.
  Fluids}  \bvol{2},  \pg{083902}.

\bibitem[Pershin {\em et~al.\/}(2019)Pershin, Beaume \& Tobias]{Pershin2019}
{\sc \au{Pershin, A.}, \au{Beaume, C.} \& \au{Tobias, S.~M.}} \yr{2019}
  \at{Dynamics of spatially localized states in transitional plane {C}ouette
  flow}.  \jt{J. Fluid Mech.}  \bvol{867},  \pg{414--437}.

\bibitem[Pringle {\em et~al.\/}(2012)Pringle, Willis \& Kerswell]{Pringle2012}
{\sc \au{Pringle, C. C.~T.}, \au{Willis, A.~P.} \& \au{Kerswell, R.~R.}}
  \yr{2012}  \at{Minimal seeds for shear flow turbulence: using nonlinear
  transient growth to touch the edge of chaos}.  \jt{J. Fluid Mech.}
  \bvol{702},  \pg{415--443}.

\bibitem[Quadrio(2011)]{Quadrio2011}
{\sc \au{Quadrio, Maurizio}} \yr{2011}  \at{Drag reduction in turbulent
  boundary layers by in-plane wall motion}.  \jt{Philos. Trans. R. Soc. A}
  \bvol{369}~(1940),  \pg{1428--1442}.

\bibitem[Quadrio \& Ricco(2004)]{Quadrio2004}
{\sc \au{Quadrio, M.} \& \au{Ricco, P.}} \yr{2004}  \at{Critical assessment of
  turbulent drag reduction through spanwise wall oscillations}.  \jt{J. Fluid
  Mech.}  \bvol{521},  \pg{251--271}.

\bibitem[Rabin {\em et~al.\/}(2014)Rabin, Caulfield \& Kerswell]{Rabin2014}
{\sc \au{Rabin, S. M.~E.}, \au{Caulfield, C.~P.} \& \au{Kerswell, R.~R.}}
  \yr{2014}  \at{Designing a more nonlinearly stable laminar flow via boundary
  manipulation}.  \jt{J. Fluid Mech.}  \bvol{738},  \pg{R1}.

\bibitem[Schmiegel \& Eckhardt(1997)]{Schmiegel1997}
{\sc \au{Schmiegel, A.} \& \au{Eckhardt, B.}} \yr{1997}  \at{Fractal stability
  border in plane {C}ouette flow}.  \jt{Phys. Rev. Lett.}  \bvol{79}~(26),
  \pg{5250}.

\bibitem[Schneider {\em et~al.\/}(2008)Schneider, Gibson, Lagha, De~Lillo \&
  Eckhardt]{Schneider2008}
{\sc \au{Schneider, T.~M.}, \au{Gibson, J.~F.}, \au{Lagha, M.}, \au{De~Lillo,
  F.} \& \au{Eckhardt, B.}} \yr{2008}  \at{Laminar-turbulent boundary in plane
  {C}ouette flow}.  \jt{Phys. Rev. E}  \bvol{78}~(3),  \pg{037301}.

\bibitem[Shi {\em et~al.\/}(2013)Shi, Avila \& Hof]{Shi2013}
{\sc \au{Shi, L.}, \au{Avila, M.} \& \au{Hof, B.}} \yr{2013}  \at{Scale
  invariance at the onset of turbulence in {C}ouette flow}.  \jt{Phys. Rev.
  Lett.}  \bvol{110}~(20),  \pg{204502}.

\bibitem[Skufca {\em et~al.\/}(2006)Skufca, Yorke \& Eckhardt]{Skufca2006}
{\sc \au{Skufca, J.~D.}, \au{Yorke, J.~A.} \& \au{Eckhardt, B.}} \yr{2006}
  \at{Edge of chaos in a parallel shear flow}.  \jt{Phys. Rev. Lett.}
  \bvol{96}~(17),  \pg{174101}.

\bibitem[Spalart {\em et~al.\/}(1991)Spalart, Moser \& Rogers]{Spalart91}
{\sc \au{Spalart, P.~R.}, \au{Moser, R.~D.} \& \au{Rogers, M.~M.}} \yr{1991}
  \at{Spectral methods for the {Navier--Stokes} equations with one infinite and
  two periodic directions}.  \jt{J. Comp. Phys.}  \bvol{96}~(2),
  \pg{297--324}.

\bibitem[Sreenivasan(1982)]{Sreenivasan1982}
{\sc \au{Sreenivasan, K.~R.}} \yr{1982}  \at{Laminarescent, relaminarizing and
  retransitional flows}.  \jt{Acta Mech.}  \bvol{44},  \pg{1--48}.

\bibitem[Stone {\em et~al.\/}(2004)Stone, Roy, Larson, Waleffe \&
  Graham]{Stone2004}
{\sc \au{Stone, P.~A.}, \au{Roy, A.}, \au{Larson, R.~G.}, \au{Waleffe, F.} \&
  \au{Graham, M.~D.}} \yr{2004}  \at{Polymer drag reduction in exact coherent
  structures of plane shear flow}.  \jt{Phys. Fluids}  \bvol{16}~(9),
  \pg{3470--3482}.

\bibitem[Stone {\em et~al.\/}(2002)Stone, Waleffe \& Graham]{Stone2002}
{\sc \au{Stone, P.~A.}, \au{Waleffe, F.} \& \au{Graham, M.~D.}} \yr{2002}
  \at{Toward a structural understanding of turbulent drag reduction: nonlinear
  coherent states in viscoelastic shear flows}.  \jt{Phys. Rev. Lett.}
  \bvol{89}~(20),  \pg{208301}.

\bibitem[Wang {\em et~al.\/}(2007)Wang, Gibson \& Waleffe]{Wang2007}
{\sc \au{Wang, J.}, \au{Gibson, J.} \& \au{Waleffe, F.}} \yr{2007}  \at{Lower
  branch coherent states in shear flows: transition and control}.  \jt{Phys.
  Rev. Lett.}  \bvol{98}~(20),  \pg{204501}.

\bibitem[Watanabe {\em et~al.\/}(2016)Watanabe, Iima \& Nishiura]{Watanabe2016}
{\sc \au{Watanabe, T.}, \au{Iima, M.} \& \au{Nishiura, Y.}} \yr{2016}  \at{A
  skeleton of collision dynamics: hierarchical network structure among
  even-symmetric steady pulses in binary fluid convection}.  \jt{SIAM J. Appl.
  Dyn. Sys.}  \bvol{15}~(2),  \pg{789--806}.

\bibitem[Zammert \& Eckhardt(2015)]{Zammert2015}
{\sc \au{Zammert, S.} \& \au{Eckhardt, B.}} \yr{2015}  \at{Crisis bifurcations
  in plane {P}oiseuille flow}.  \jt{Phys. Rev. E}  \bvol{91},  \pg{041003(R)}.

\end{thebibliography}

\end{document}